\documentclass[aps,prd,twocolumn]{revtex4-1}

\usepackage{aas_macros}
\usepackage{epsfig}
\usepackage{latexsym}
\usepackage{times}

\def\i{\item}

\newcommand{\bed}{\begin{displaymath}}
\newcommand{\eed}{\end{displaymath}}
\newcommand{\bef}{\begin{figure}}
\newcommand{\eef}{\end{figure}}
\newcommand{\ben}{\begin{enumerate}}
\newcommand{\bei}{\begin{itemize}}
\newcommand{\eei}{\end{itemize}}
\newcommand{\een}{\end{enumerate}}
\newcommand{\beq}{\begin{equation}}
\newcommand{\eeq}{\end{equation}}
\newcommand{\ber}{\begin{eqnarray}}
\newcommand{\eer}{\end{eqnarray}}

\newcommand{\bb}{\bf B}
\newcommand{\fb}{\bf f}

\newcommand{\jb}{\bf J}
\newcommand{\nb}{\bf \nabla}
\newcommand{\rb}{\bf r}

\newcommand{\msun}{\mbox{{\rm M}$_{\odot}$}}

\newcommand{\gsim}{\raisebox{-0.3ex}{\mbox{$\stackrel{>}{_\sim} \,$}}}

\usepackage{xcolor}
\newcounter{attnctr} 

\usepackage{ulem}

\begin{document}

\title{Strong constraints on magnetized white dwarfs surpassing the 
       Chandrasekhar mass limit}
\author{Rajaram Nityananda}
\affiliation{Indian Institute of Science Education and Research, Pune 
             411008, India}
\email[]{rajaram.nityananda@iiserpune.ac.in\\
         currently at Azim Premji University, Bangalore 560100, India}
\author{Sushan Konar}
\affiliation{National Centre for Radio Astronomy, Tata Institute of 
             Fundamental Research, Pune 411007, India}
\email[]{sushan@ncra.tifr.res.in}
\homepage[]{http://www.ncra.tifr.res.in:8081/~sushan/}

\begin{abstract}
We show that recently proposed  white dwarf models with masses well in
excess of the Chandrasekhar limit,  based on modifying the equation of
state by  a super-strong  magnetic field in  the centre, are  very far
from equilibrium because  of the neglect of Lorentz  forces.  An upper
bound  on the  central magnetic  fields, from  a  spherically averaged
hydrostatic equation, is much smaller than the values assumed.  Robust
estimates  of  the  Lorentz  forces  are also  made  without  assuming
spherical averaging.  These again bear out the results obtained from a
spherically averaged  model. In  our assessment, these  estimates rule
out the  possibility that magnetic tension could  change the situation
in  favor   of  larger  magnetic  fields.   We   conclude  that  such
super-Chandrasekhar  models are  unphysical and  exploration  of their
astrophysical consequences is premature.
\end{abstract}

\pacs{}


\maketitle

\date{\today}

\section{Introduction}

Models  for  `white dwarf' like  stars  (i.e  stars supported  against
gravity  by electron  degeneracy pressure)  with  masses significantly
exceeding  the Chandrasekhar limit  (e.g 2.3  - 2.6~\msun),  and radii
significantly smaller  than hitherto considered possible  ($\sim$ 70 -
600~km),   have   been   proposed   \cite{das12a,das12b}   and   their
astrophysical   consequences  explored   \cite{das13a,das13b}.   These
models  are based on  the altered  equation of  state coming  from the
quantization  of electron motion  in super-strong  ($\gsim 10^{15}$~G)
magnetic  fields.   We  show  below  that  these  models  are  not  in
hydrostatic equilibrium, a fact  missed in the original and subsequent
work which  ignores the unavoidable  gradient of magnetic  pressure. A
brief comment to this effect has been submitted \cite{nitya14} and the
present  paper  gives  more  details  and, in  particular,  lifts  the
assumption of spherical symmetry.

The main concern of this note  is a counter-intuitive feature of these
models  --  the magnetic  pressure  ($P_m$)  is  not included  in  the
equation  of hydrostatic  equilibrium  even though  its  value at  the
centre exceeds the  electron pressure ($P_e$) by nearly  two orders of
magnitude.   Given  that the  models  balance  gravity solely  against
electron pressure, this implies that the field pressure/energy density
greatly  exceeds  the  traditional  dimensional  estimate  of  central
pressure, given by $G M^2/R^4$, where  $M$ and $R$ denote the the mass
and the  radius of  the spherical  star and  $G$ is  the gravitational
constant.   The  justification  given   (in  \cite{das12a},  and  most
recently in \cite{das13c}) is that the field is uniform in the central
region of  interest, and therefore there  is no force coming  from the
gradient of  $P_m$. The field  is then presumed  to taper off  to much
smaller values  at the surface (consistent  with observations) without
affecting the analysis  which relies solely on  electron pressure.  In
the next section, we show in a sphercially averaged model that this is
not possible  -- the equation  of equilibrium  is violated by  a large
factor  in the  transition region,  where the  strong uniform  central
field reduces  to much  smaller values.   In Sec. II  B, we  relax the
assumption of spherical averaging, and use the magnetic virial theorem
to derive  general bounds on  the central  fields which are  still far
less that the  proposed values.  In Sec. III we  comment upon the need
to include  general relativity  and other  effects while  dealing with
highly relativistic electrons in extremely strong magnetic fields.

\section{Bounds on the central magnetic field}
\subsection{Spherical symmetry} 

We initially  restrict to an averaged spherically  symmetric model, as
in \cite{das12a}, in which the stress tensor of the magnetic field can
be replaced by  an equivalent isotropic pressure.  We  take this $P_m$
to be $B^2/24  \pi$, one third the trace of  the Maxwell stress tensor
(also one third the energy density), where $B$ is the magnitude of the
magnetic field.  Then at a radius $r$ inside the star, the equation of
hydrostatic equilibrium reads :
\beq
\frac{d P_e}{dr} + \frac{d P_m}{dr} = - \rho(r) g(r), 
\label{eq-hydro1}
\eeq
where $g(r) (=G M(r)/r^2)$  is the radially inward gravitational force
on a unit mass, $\rho(r)$ and $M(r)$ being the mass density at and the
total mass  contained within, a  radius $r$.  In the  proposed models,
the second  term on  the left  hand side is  assumed to  be negligible
compared    to    the    first.     Integrating    both    sides    of
Eq.(\ref{eq-hydro1}), from  the centre ($r=0$)  to the surface  of the
star ($r=R$), denoted by suffixes $c$ and $s$, we obtain
\beq
(P_{\rm ec} - P_{\rm es}) +  (P_{\rm mc} - P_{\rm ms}) = \int^R_0 \rho
(r) g(r) dr.
\label{eq-press}
\eeq
Clearly, the second  bracket on the right hand  side should be smaller
than the  first, if  the neglect of  the magnetic contribution  to the
hydrostatic  equation  is to  be  valid.   But  (equating the  surface
pressures to zero) the exact  opposite is true of the proposed models,
the  magnetic  pressure being  very  much  greater  than that  of  the
degenerate electrons, which is an obvious contradiction.

\bef  
\epsfig{file=fig01.ps,width=145pt,angle=-90}
\caption[]{Radial density profiles. The solid line denotes the actual
  profile as calculated by Das \& Mukhopadhyay (2012a) for a star with
  $B_c = 8.81 \times 10^{15}$~G, and  $E_{\rm F max} = 20 m_e c^2$ and
  the dotted line is a $\sin(x)/x$ curve, where $x$ is equal 
  to $\pi r/r_{\rm max}$.}
\label{fig01}  
\eef 

We consider a particular  case explored in \cite{das12a} to illustrate
this  point.  Fig.\ref{fig01} shows  the radial  density profile  of a
proposed stellar  model which  has a central  magnetic field  equal to
$8.81 \times  10^{15}$~G. The maximum Fermi energy  of the constituent
electrons is  assumed to be $20 m_e  c^2$, where $m_e$ is  the mass of
the electrons  and $c$ is  the velocity of  light.  We note  that this
density  profile  is well  approximated  by  a  function of  the  form
$\sin(x)/x$.   This, of  course, is  the  form of  the radial  density
profile for  a star which is  a $n=1$ polytrope. For  a polytrope, one
assumes the gas pressure to be given by
\beq
P = K \rho^\gamma = K \rho^{(n +1)/n}, 
\label{eq-polyt}
\eeq
where $\gamma$ is the adiabatic index and $n$ is called the polytropic
index.  Here,  $K$ is a  dimensional constant characterizing  the gas.
Physically, the $n=1$ polytrope corresponds  to the case of an extreme
relativistic  gas with  unfilled lowest  Landau level.   It  should be
noted that non-magnetic white dwarfs are described by $n = 1.5$ in the
region where electrons are non-relativistic,  rising to $n = 3$ as the
electrons become relativistic.

In  Fig.\ref{fig02} we  compare  the radial  variations  of $P_e$  and
$P_m$.  To calculate $P_m$ we  assume two different radial profiles of
the magnetic field,  in both cases matching to a  value of $10^9$~G at
the surface, the maximum observed surface field for white dwarfs.
 
\bef
\epsfig{file=fig02.ps,width=145pt,angle=-90}
\caption[]{Radial pressure  profiles.  The solid  line ($P_e$) denotes
  the  pressure due  to  the  electrons in  presence  of a  quantizing
  magnetic field.  The dotted  ($P^1_B$) and the dash-dotted ($P^2_B$)
  lines represent the  pressure due to the magnetic  field.  A central
  field  of $8.8 \times  10^{15}$~G has  been assumed  to fall  off to
  $10^9$~G at  the surface. While  the dash-dotted curve  represents a
  linear  fall-off, the  dotted curve  shows the  case of  a power-law
  fall-off for the magnetic field. }
\label{fig02} 
\eef 
 
This figure confirms our earlier  assertion. No attempt to taper off a
large,  uniform  magnetic field  in  the  centre  to essentially  zero
($B_s/B_c <<  1$) at the surface can  avoid a gradient of  $P_m$ to be
much  larger than  that  of  $P_e$ (at  least  in certain  locations),
which balances gravity in the proposed models.

\subsection{Beyond  spherical symmetry}

Our  constraints on the  equilibrium of  models supported  by electron
pressure alone,  have so far been  derived in the  spirit of spherical
averaging.  This  would be strictly  applicable only if the  field was
sufficiently  disordered to  result in  an average  isotropic pressure
within a region smaller than  the scale length over which pressure and
density vary significantly. It has been pointed out~\cite{das13c} that
magnetic  tension in  an ordered  field has  been left  out of  such a
picture.  According to this  view, the magnetic tension could actually
act  in an  opposite way  to magnetic  pressure, and  possibly  play a
significant role  in stabilizing a non-spherical  configuration with a
suitably ordered field.

We  first examine  this possibility  by deriving  a constraint  on the
field strength in  the case of a poloidal field.   Assume the field in
the centre  to be  along the $z$-axis.  The model is  now axisymmetric
rather than spherical, with  the shorter dimension along the $z$-axis,
as  expected  from the  steeper  density  gradient  needed to  balance
gravity  aided  by  magnetic  tension.   However,  the  tension  along
$z$-axis is now  accompanied by a lateral pressure  $P_{m \perp}$ ($ =
B^2/8 \pi$) in the $xy$-plane.   We assume the fraction of the central
flux  leaking out  of the  star to  be very  small, since  the maximum
surface fields  observed in white  dwarfs are six orders  of magnitude
smaller than the  central fields being envisaged.  In  this case, most
of the field lines would necessarily have to return with opposite sign
and  cross the  equatorial plane  within the  star.  The  situation is
shown in Fig.\ref{fig03}, for a poloidal field configuration.

We can now apply our earlier  argument in the equatorial plane, with a
three times stronger magnetic  pressure at the centre ($P_{m \perp}$),
and an appropriately weaker gravity term.  The gravitational potential
gradient  term  in  the  hydrostatic  equation  gets  reduced  in  the
equatorial plane due to oblateness.   Therefore it appears that in the
poloidal case, equilibrium in the $z$-direction would have to be bought
at the price of even greater disequilibrium in the equatorial plane.

\bef 
\epsfig{file=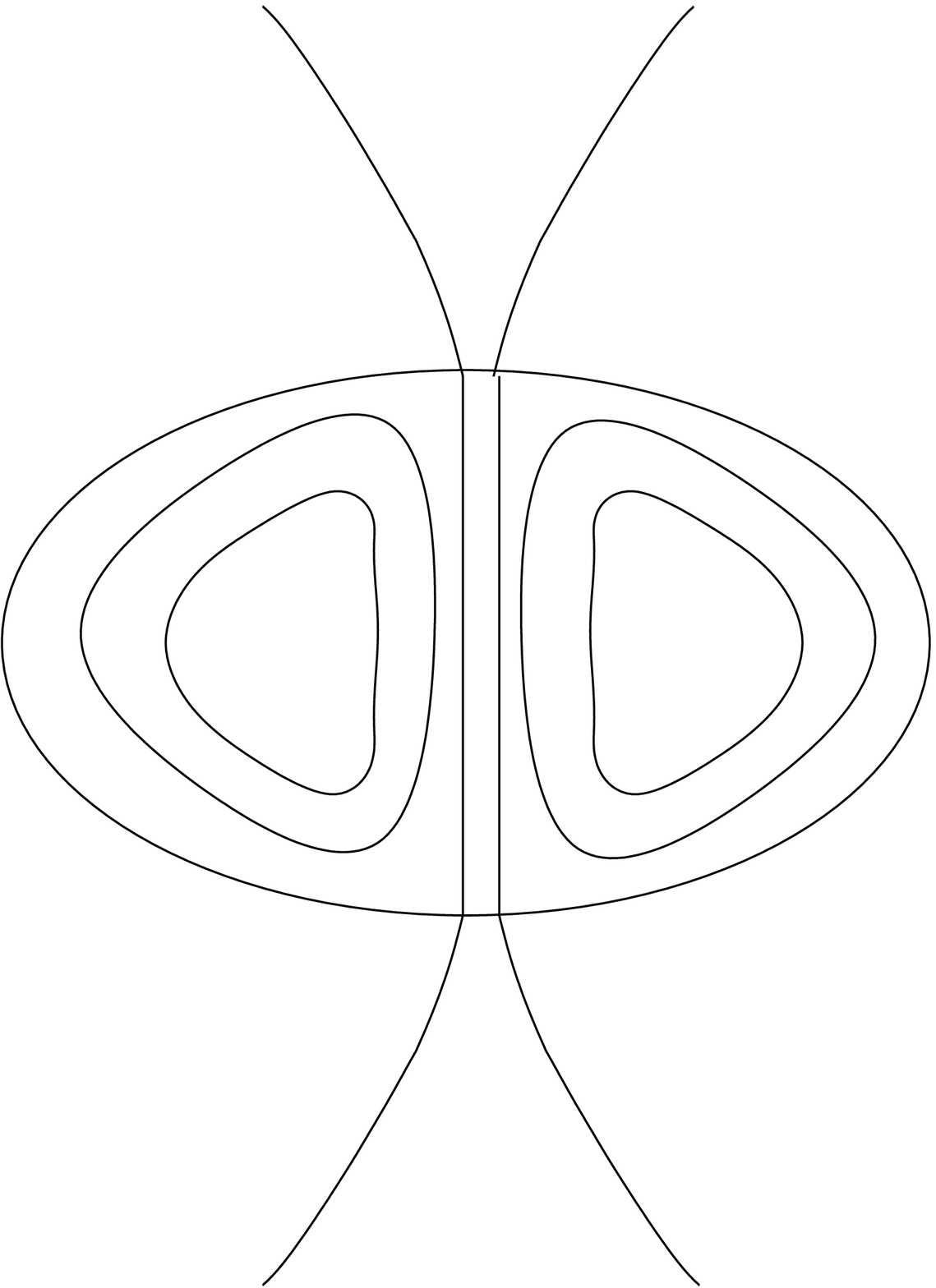,width=145pt,angle=0}
\caption[]{Schematic  representation  of  a  poloidal  magnetic  field
  inside a  self-gravitating oblate spheroid. Notice that  most of the
  field lines close inside the  star.  This is expected in a situation
  where there is a strong field in the central region tapering down to
  a very small value near the surface.}
\label{fig03}
\eef
 
However, one has to consider the possibility of a toroidal field whose
tension could  help maintain equilibrium in the  equatorial plane (and
stabilize the poloidal  configuration as well). But now  this would be
attained at the cost of outward forces away from the equatorial plane.
We  constrain this more  general situation  below, using  the magnetic
virial theorem.

The  Lorentz force  density $\fb^L$  inside a  continuous  medium with
current density $\jb$ and magnetic field $\bb$ is given by,
\beq
\fb^L = \frac{\jb \times \bb}{c}
      = \frac{1}{4 \pi} (\nb \times \bb) \times \bb.
\label{eq-loren1}
\eeq
By  the  virial  identity  for  the Lorentz  force  (for  details  see
\cite{chand81}, page 158, Eq.78) we have
\beq
\int^R_0 {\rb.\fb^L} \, d^3 {\rb}
= \int^R_0 {\frac{\bb^2}{8\pi}} d^3 {\rb} = E_{\rm B},
\label{eq-loren2}
\eeq
where $E_{\rm B}$ is the total magnetic energy of the system and $\rb$
is the radius  vector.  In writing this we  have neglected the surface
terms at the upper limit  of integration $R$, the stellar radius. This
is justified in view of the surface fields being much smaller than the
postulated  central  fields.    One  immediate  conclusion  from  this
identity is that  the average value of $\rb.\fb^L$  is positive.  This
implies  that   the  average   Lorentz  force  is   outwards,  tension
notwithstanding (this  is in conformity with  the spherically averaged
model which has a positive isotropic pressure).

Since the  outward magnetic force  cannot exceed gravity  anywhere, we
proceed as follows.  We use the  virial identity to give a lower bound
on the  maximum value of  the Lorentz force  density, in terms  of the
central magnetic field.  This has to be less than the maximum value of
the  gravitational force density.   The resulting  upper bound  on the
central field is conservative but will  be enough to rule out the kind
of  central  fields being  postulated.   The  method  is general,  but
is illustrated for polytropic models below.

To  obtain  the  maximum  allowed  magnitude  of  $\bb$,  we  make  an
underestimate of the maximum value  of the Lorentz force density ${\bf
  f^L}$ (as a function of radius) within the star. To this end, we use
Eq.(\ref{eq-loren2}) in the following form,
\ber
f^L_{\rm max} \int^R_0 \frac{{\rb.\fb^L}}{f^L_{\rm max}} \, d^3 {\rb}
= E_{\rm B}, \, \, \, \Rightarrow \, \, \,
f^L_{\rm max} = \frac{E_{\rm B}}{I}.
\label{eq-loren3}
\eer
where,  $f^L_{\rm max}$  is the  maximum  value of  the Lorentz  force
density   and    $I$   is   the   integral    defined   as   $\int^R_0
{\frac{\rb.\fb^L}{f^L_{\rm  max}}}   \,  d^3  {\rb}$.    To  obtain  a
conservative  lower limit  of  $f^L_{\rm max}$,  corresponding to  the
conservative lower  limit of  the maximum magnetic  field, we  need to
underestimate  $E_{\rm B}$  and overestimate  $I$.  A  lower  bound on
$f^L_{\rm max}$ is therefore given by,
\beq
f^L_{\rm max} > \frac{E_-}{I_+},
\label{eq-loren4}
\eeq
where  $E_-$ is  an  underestimate of  $E_{\rm  B}$, and  $I_+$ is  an
overestimate of $I$. For an equilibrium model, $f^L_{\rm max}$ is less
than the maximum value of the gravity term.

We  now  examine  the  maximum  value  of  the  gravity  term  in  the
hydrostatic equation by considering the inward gravitational force per
unit volume, $f^g(r) \, (= \rho(r)  g(r))$. For a given total mass and
radius, the value  and the location of $f^g_{\rm  max}(r)$ is strongly
dependent on  the central concentration of the  mass distribution, and
less  so on  the flattening  so long  as it  is modest.   We therefore
illustrate this  by means of  polytropic spherical models,  though the
method  is general.   Fig.\ref{fig04}  shows the  radial variation  of
$f^g(r)$  for different polytropic  models.  It  is observed  that the
gravitational force density increases from zero at the centre, reaches
a maximum value at a certain radius ($R^g_m$) and then gradually falls
to  zero   again  at  the  surface.   For   a  centrally  concentrated
$n=3$~polytropic model it peaks early ($R^g_m \sim 0.2 R_*$) while for
$n=1$  model it  peaks at  $R^g_m \sim  0.5 R_*$,  where $R_*$  is the
corresponding stellar radius.

\bef 
\epsfig{file=fig04.ps,width=145pt,angle=-90}
\caption[]{$f^g \, (= \rho g)$ vs.  fractional radius $x \, (= r/R_*)$
  for different polytropic ($n =  1, 3$) models; $f^g(r)$ for each $n$
  being scaled  to unity at the  maximum for ease  of comparison. Note
  that  the $n=3$  case  corresponds  to a  star  supported by  highly
  relativistic particles.  }
\label{fig04}
\eef 

For a star  to be in equilibrium it is  necessary that $f^L_{\rm max}$
should  be smaller  than $f^g_{\rm  max}$. We  have seen  that Lorentz
force scales as the gradient  of the field.  Therefore, $f^L_{\rm max}
<  f^g_{\rm max}$  implies  that  the uniform  magnetic  field at  the
centre,  which has  zero Lorentz  force, would  also have  to  drop to
smaller values around the location of maximum gravity.  This is indeed
the  most  favorable  situation  for equilibrium.   If  Lorentz  force
exceeds gravity there, it would do so even more if the falloff were to
occur at  some other  $r$, greater or  less than $R^g_m$.   Using this
physical idea we  set $B(r) = B_c$ for $r <  R^g_m$. Then the magnetic
energy is given by,
\beq
E_B = \int_0^{R^g_m} \frac{B_c^2}{8 \pi} d^3 {\rb}
      + \int_{R^g_m}^R \frac{B_{nc}^2}{8 \pi} d^3 {\rb}
      \geq \frac{4 \pi}{3} (R^g_m)^3 \frac{B_c^2}{8 \pi},
\label{eq-eB}
\eeq since $B_{nc}$, the 'non-central' magnetic field for $r < R^g_m$,
is always smaller than the  uniform central field $B_c$. Therefore, we
take the underestimate $E_-$ of the field energy, $E_B$, to be,
\beq
E_- = \frac{1}{6} (R^g_m)^3 B_c^2.
\label{eq-emin}
\eeq

To obtain  an overestimate  of $I$  we note that  the integrand  is $r
(f^L/f^L_{\rm max})  \cos \theta$, where both  $f^L/f^L_{\rm max}$ and
$\cos \theta$  are always  less than unity.   Then an  overestimate is
obtained by replacing these two factors by unity we have,
\beq
I = \int^R_0 {\frac{\rb.\fb^L}{f^L_{\rm max}}} \, d^3 {\rb}
    \leq f \int^R \, \cos \theta \, r \, d^3 {\rb}
  = \pi R^4
  = I_+ \,.
\label{eq-imax}
\eeq

Requiring  $f^L_{\rm max}$  to be  less than  $f^g_{\rm max}$  we then
obtain   a  constraint  on   $B_c$  by   using  Eq.s(\ref{eq-loren4}),
(\ref{eq-emin})  \&  (\ref{eq-imax}).    Expressed  in  terms  of  the
equivalent isotropic  magnetic pressure  at the centre,  $P_{cm}$, the
result is
\ber
P_{cm} < \frac{1}{4} \, (\rho \, g)_{max}
     \, \left(\frac{R^g_m}{R}\right)^{-3} \, R,
\label{eq-pcm}
\eer
taking $P_m  = B^2/24 \pi$ as  before.  In order to  express the right
hand side in  terms of average stellar quantities, we  shall now use a
spherically  symmetric   polytropic  model  purely   as  a  convenient
parametrization  of  a  family   of  models  with  increasing  central
concentration.

To  facilitate easy  understanding of  what follows,  we lay  down the
basic structure of  the polytropic models here~\cite{chand39,clayt68}.
The solution for the structure  of a gravitationally bound object with
a  polytropic   equation  of  state  depends  upon   the  coupling  of
Eq.(\ref{eq-polyt}) to the  condition of hydrostatic equilibrium given
by Eq.(\ref{eq-hydro1}). From this, it follows that,
\beq
\frac{1}{r} \frac{d}{dr}\left(\frac{r^2}{\rho}\frac{dP}{dr}\right)
= - 4 \pi G \rho,
\label{eq-hydro2}
\eeq
where $P$ denotes the total pressure  at a radius $r$ inside the star.
Motivated by the  fact that the density is  proportional to $T^n$ ($T$
is  the  system temperature)  in  a polytropic  gas  of  index $n$,  a
convenient definition of $\rho$ is,
\beq
\rho = \lambda \theta^n,
\eeq
where $\lambda$ is a constant.  Substitution of the values of pressure
and  density for a  polytrope of  index $n$  into Eq.(\ref{eq-hydro2})
gives us,
\beq
(n+1) K \lambda^{1/n} \frac{1}{r^2} \frac{d}{dr} 
\left(r^2 \frac{d\theta}{dr}\right) 
= - 4 \pi G \lambda \theta^n .
\eeq
This reduces to
\beq
\frac{1}{\xi^2} \frac{d}{d\xi} 
\left( \xi^2 \frac{d\theta}{d\xi}\right) = - \theta^n,
\label{eq-lane}
\eeq
by defining a dimensionless distance variable $\xi = r/a$, where
\beq
a = \left[ \frac{(n+1) K \lambda^{(1-n)/n}}{4\pi G}\right]^{\frac{1}{2}}.
\eeq
Eq.(\ref{eq-lane}) is called the Lane-Emden equation for the structure
of a polytrope of index $n$.

Let us  now consider the central gravitational  pressure, $P_c$, which
is the pressure  implied by the equation of  equilibrium, and the mass
and  radius,  for  a  given  degree  of  central  concentration  which
increases with the polytropic index  $n$.  It is precisely $P_c$ which
is  exceeded by  the central  magnetic  pressure in  the models  under
discussion.   

Using  the   standard  solutions  for  various   quantities  inside  a
polytropic star, we obtain the following relation between $P_{\rm cm}$
and $P_c$ :
\beq
P_{cm} 
< - \frac{1}{4} \, (n+1) \, 
    \left(\xi \, \theta^n \theta' \right)_{R^g_m} \,
    \left(\frac{R^g_m}{R}\right)^{-4} \,
    \times P_c.
\label{eqn10}
\eeq
%
To obtain the above condition we have used the following standard relations,
generic to any polytropic model,
\ber
\rho_c &=& - \frac{1}{4 \pi} \frac{\xi(R)}{\theta'(R)} \, \frac{M}{R^3}, \\
P_c &=& \frac{1}{4 \pi (n+1) (\theta'(R))^2} \, \frac{G M^2}{R^4}, \\
M(r) &=& \frac{r^2 \theta'(r)}{{R}^2 \theta'(R)} \, M;
\label{eqn11-13}
\eer
where $\theta$ is the Lane-Emden  function of polytropic order $n$ and
$\xi$   is   the    dimension-less   radial   parameter   defined  
above. 
Here $\theta'$ denotes the derivative of $\theta$ with respect
to $\xi$. $\rho_c$, $M$ and $R$ stand for the central density,
the total mass and the radius of the star respectively.

Evidently, the ratio between $P_{\rm cm}$ and $P_c$, given by,
\ber
\mathcal{Q}(n) 
= - \frac{1}{4} \, (n+1) \, 
    \left(\xi \, \theta^n \theta' \right)_{R^g_m} \,
    \left(\frac{R^g_m}{R}\right)^{-4} \,
\label{eqn14}
\eer
now depends solely on the  polytropic index $n$. 

\begin{center}
\begin{table} 
\begin{tabular}{|c|c|c|c|c|} \hline
&&&& \\
n & 1 & 1.5 & 2 & 3 \\ 
&&&& \\ \hline
$\mathcal{Q} \left(\frac{R^g_m}{R}\right)^4$ & 2.10 & 1.89 & 1.76 & 1.62 \\ 
&&&& \\ \hline
\end{tabular}
\caption[]{Variation of the numerical factor $\mathcal{Q}$ with
polytropic index $n$.}
\label{tab01}
\end{table}
\end{center}

It  is seen  from table-\ref{tab01}  that  for a  range of  polytropic
indices the  proportionality between  the conservative upper  limit on
$P_m$ and  $P_c$ is  dependent upon the  location of  $f^g_{\rm max}$,
apart from a small factor  (which, in polytropic models depend only on
$n$).  This result itself  is remarkably insensitive to the polytropic
index, thanks  to the  choice of $R^g_m$  as the  independent variable
though the actual  region of high gravity and  magnetic field gradient
naturally moves inwards as $n$ increases.

There is  some scope  for weakening the  bound by invoking  factors we
have  not included.  These  are non  sphericity, and  a non-polytropic
running of pressure and density.  The upper limits on the integrals in
$E_-$  and $I_+$  could  differ from  $R^g_m$  though not  by a  large
factor.  It should now be amply clear, with due allowance for magnetic
tension,  the central  magnetic pressure  cannot exceed  that inferred
from the mass distribution by a factor of 100 or larger.

Consider, for example,  the extreme model which has  $M = 2.58$~\msun,
$R  = 69.5$~KM  with a  central magnetic  field of  $B_c =  8.8 \times
10^{17}$~G~\cite{das13a}.   Since the  equation of  state  matches the
$n=1$  polytropic case  very closely,  using  the above  table we  can
obtain the upper bound to the  central magnetic field.  It can be seen
from Fig.[\ref{fig04}] above  that $R^g_m/R$ is close to  0.5 for this
case. Using  this we  find the  maximum central field  to be  given by
$B_{\rm upper-bound} \simeq 10^{16} $~G.  This is almost two orders of
magnitude smaller than  the field claimed to be  present in the centre
of such an object.

Although we  have used spherically symmetric  polytropes to illustrate
the  trend with varying  central concentration,  the argument  is more
general which can be used as  a reality check given the running of density
and magnetic field in any proposed model, even an anisotropic one.  We
conclude  that   ordered  fields   in  an  anisotropic   model  cannot
qualitatively change the conclusions  drawn from the average spherical
model.

\section{The extreme relativistic limit}  

The electrons, in the  models under consideration, have Fermi energies
significantly above  their rest energy.  We point  out certain generic
features of such extreme relativistic systems.  Consider the case when
the magnetic field  is such that the lowest Landau  level (LLL) is just 
full.  Then the relation between the electron density ($n_e$) and the
the cyclotron frequency  ($\omega_c = eB/m_ec$) is~\cite{lai91} (using
$\hbar = m_e = c = 1$ from here onward),
\beq
n_e=\omega_c^{3/2}/\sqrt{2}\pi^2.   
%
\label{eqn16}
%
\eeq
where $p_F$  is the  Fermi momentum of  the electrons.   Therefore the
electron pressure is given by,
\beq
P_e = \frac{1}{2} n_e E_F = \frac{1}{\sqrt{2} \pi} \omega_c^2, 
\label{eqn16}
\eeq
where $E_F$ is  the Fermi energy of the  electrons and is proportional
to   $n_e$   for  ultra-relativistic   particles.    From  these   two
expressions we note the following results.
\ben 
\i  At the  centre of  the  star, with  an exact  LLL filling  $P_{\rm
  c-LLL}$  is, proportional to  the four-third  power of  the electron
density,  just  as  $P_{\rm  eC}$,  the  gas  pressure  in  the  usual
(Chandrasekhar) case.   But it has a  different numerical coefficient,
that  is $P_{\rm  e-LLL}  =  2^{-1/3} \,  \pi^{2/3}  \, n_e^{4/3}$  as
opposed to $P_{\rm eC} = 3^{1/3} \, 4^{-1} \, \pi^{2/3} \, n_e^{4/3}$.
Therefore  $P_{\rm e-LLL}$  is  a  factor of  $4  \times 6^{-1/3}$  or
approximately  2.2 times  greater  than $P_{\rm  eC}$.   But at  lower
densities, as  we move  away from the  centre, the pressure  varies as
$n_e^2$, so  the running of density  with radius is very  close to the
n=1 polytrope, which is a  stiffer equation of state. A combination of
the enhanced numerical factor and  a stiffer polytropic index (for the
equation  of state)  is then  responsible for  the super-Chandrasekhar
mass           obtained            in           the           proposed
models~\cite{das12a,das12b,das13a,das13b},  of course with  neglect of
magnetic pressure.
\i  Plasma beta  ($\beta_p$), the  ratio of  gas pressure  to magnetic
pressure (taken, as  before, to be $B^2/24\pi$), at  the LLL condition
is  independent of  the  field strength  in  the extreme  relativistic
limit,  and  is given  by  $12  \alpha/\pi$,  i.e around  $2.8  \times
10^{-2}$  ($\alpha=e^2/c\hbar$   is  the  fine   structure  constant).
Generically, the magnetic pressure  is two orders of magnitude greater
than the  electron pressure  in a relativistic  model with only  a few
Landau levels filled. This  incidentally implies, if the electrons are
already relativistic,  then the rest  energy of the magnetic  field is
rather significant. Because  a hundred times $20 m_e  c^2$ is not very
far from  $4000 m_e c^2$ which is  the rest energy of  the nuclei, per
electron.  This factor goes in the direction of softening the equation
of  state, towards  $p \propto  \rho$, and  points to  the need  for a
general  relativistic treatment since  pressure gravitates  in general
relativity.   It  should  be  noted  that  for  the  particular  model
mentioned above~\cite{das13a} a field of $\sim 10^{17}$~G would have a
rest  mass density  comparable to  the  average density  of the  star.
Since higher fields are being  postulated in the central regions it is
clear that general relativistic effects would play a major role.

Another  important factor  which would  modify the  equation  of state
under such  conditions is neutronisation  (inverse $\beta$-decay), the
absorption of electrons by  protons to produce neutrons, which becomes
favorable with  increasing electron energy~\cite{lai91}.   As a result
of  neutronisation  the  electron   number  decreases  and  the  ionic
component of the pressure (which has been completely neglected so far)
begins   to   become  important.    Moreover,   at  higher   densities
pycnonuclear  reactions would  modify  the composition  of the  matter
making  it even  more  difficult for  the  star to  be  treated as  an
ordinary  white  dwarf  (with  standard compositions).   The  proposed
formalism does not provide for these energetically favorable processes
either,    the    effect    of    which    has    been    investigated
recently~\cite{chame13}. This particular issue  (and a number of other
concerns)   is    now   being    addressed   by   other    groups   as
well~\cite{coelh13}.  The results from a self-consistent investigation
into  the structures  of  strongly magnetized  white  dwarfs has  just
become available~\cite{bera14} and it is seen that while the masses of
such objects do exceed the traditional Chandrasekhar limit, neither do
the structures deviate too far  from those of the non-magnetized white
dwarfs nor do the maximum  field strengths ($B \sim 10^{14}$~G) differ
significantly  from  those expected  from  simple stability  arguments
presented here.
\een
 
\section{Conclusions}  

We would  like to reiterate that  the effects we  are considering here
arise  due to a  rather basic  physical reason  -- the  currents which
generate the magnetic field flow somewhere inside the star (one is not
considering stars in external fields!) and they must experience a $\jb
\times \bb$  force.  There exist {\em force  free} configurations with
current flowing parallel to $\bb$, which have been extensively studied
in  the context  of solar  physics, for  example.  But  it is  a known
characteristic   of   these  configurations   that   the  forces   are
redistributed to the boundaries  rather than vanishing everywhere.  In
the words of one of the founders  of the subject, `` {\em while we may
  be  able to cancel  the stresses  inside a  given region,  we cannot
  arrange for  its cancellation everywhere}''  (\cite{chand81}, p-159).
Our basic point  is that it is critical to  account for the transition
from  the force  free, uniform  field in  the centre  to  much smaller
fields outside,  via a region  of strong average outward  forces which
carries the current.  The transition zone could be narrow, for example
if we had a spherical region  with a uniform field, and a dipole field
outside, both have  zero curl and are in current  free and hence force
free regions~\cite{das13c}.  However, all the current  is then carried
in  the boundary in  between these  regions, and  the integral  of the
force  density,  continues to  be  finite,  from  the virial  theorem.
Models  which do  not  account  for these  considerations  are not  in
equilibrium  and are  clearly unphysical.   To conclude  then,  in our
view, it is quite premature to construct astrophysical scenarios until
at least one  equilibrium model of a star has  been obtained with such
extreme conditions. \\

\begin{acknowledgments}

We  thank  Dipankar Bhattacharya  for  sharing  his  insights in  many
discussions   of  this   problem.   We   appreciate   and  acknowledge
correspondence  and discussions  with Banibrata  Mukhopadhyay, carried
out in the face of our widely differing viewpoints.

\end{acknowledgments}


\bibliography{adsrefs}

\end{document}


\title{Erratum : Strong constraints on magnetized white dwarfs surpassing the 
        Chandrasekhar mass limit [Phys. Rev. D 89, 3017 (2014)]}
\author{Rajaram Nityananda}
\affiliation{Azim Premji University, Bangalore 560100, India}
\email[]{rajaram.nityananda@apu.edu.in}
\author{Sushan Konar}
\affiliation{National Centre for Radio Astronomy - Tata Institute of 
             Fundamental Research, Pune 411007, India}
\email[]{sushan@ncra.tifr.res.in}
\homepage[]{http://www.ncra.tifr.res.in:8081/~sushan/}

\begin{abstract}
We correct certain numerical  estimates made in our paper. This
correction  however does not  alter the  conclusion that  when Lorentz
forces are  taken into account  the super-massive white  dwarf models,
obtained simply  by modifying the equation  of state in  presence of a
super-strong magnetic field in the centre, fail to be in equilibrium.
\end{abstract}

%

\maketitle

\date{\today}

\begin{center}
\begin{table} 
\begin{tabular}{|c|c|c|c|c|} \hline
&&&& \\
n & 1 & 1.5 & 2 & 3 \\ 
&&&& \\ \hline
&&&& \\ 
$\mathcal{Q} \left(\frac{R^g_m}{R}\right)^4$ & 6.31 & 16.31 & 53.86 & 662.84 \\ 
&&&& \\ \hline
\end{tabular}
\caption[]{Variation of the numerical factor $\mathcal{Q}$ with
polytropic index $n$.}
\label{tab01}
\end{table}
\end{center}

The ratio  between $P_{\rm cm}$ (upper bound to the central magnetic  pressure) 
and $P_{\rm c}$ (central  pressure  of  the  white  dwarf) required  for  the  minimal
stability of the star is given by,
\ber
P_{\rm cm} / P_{\rm c} = 
\mathcal{Q}(n) = 
-  \frac{1}{4}  \, (n+1)  \,  \left(\xi \,  \theta^n
\theta' \right)_{R^g_m} \, \left(\frac{R^g_m}{R}\right)^{-4} \,
\label{eqn14}
\eer
as stated  in Equation (18) of our  original work~\cite{nitya14}. The
numerical values of $\mathcal{Q}$  (dependent solely on the polytropic
index $n$) provided in Table  I of \cite{nitya14} were incorrect. The
corrected values are given above.

Our earlier  statement below the  table that the ratio  $\mathcal{Q} =
P_{\rm  cm}/P_{\rm c}$  is  relatively insensitive  to the  polytropic
index  $n$   is   not   correct   -   that    statement   applies   to
$\mathcal{Q}(R_m/R)^4$.  The consequence  of this  correction  for the
model  which  has $M  =  2.58$~\msun, $R  =  69.5$~KM  with a  central
magnetic  field of  $Bc  = 8.8  \times  10^{17}$~G \cite{das13a}, is  as
follows.  We  use  $n=1$ which  is  close  to  the situation  in  this
model. The upper  bound to the central magnetic field  comes out to be
$3.4 \times 10^{17}$~G  and not $10^{16}$~G as stated  in our original
paper~\cite{nitya14}.  We  note that  this  still  implies a  central
magnetic  pressure in the  proposed model  nearly seven  times greater
than our conservative upper bound. Our main conclusion that the models
neglecting Lorentz forces cannot be in equilibrium stands.

We thank U. Das and B. Mukhopadhyay for pointing our the numerical errors in
our work~\cite{nitya14}.


\bibliography{adsrefs}